\documentclass[10pt,twoside]{hsqcd}
\usepackage{epsf,amsmath}
\usepackage{graphicx}

\newcommand{\bp}{\mbox{\boldmath $p$}}

\newcommand{\br}{\mbox{\boldmath $r$}}

\newcommand{\bb}{\mbox{\boldmath $b$}}

\newcommand{\bC}{\mbox{\boldmath $C$}}
\newcommand{\ket}[1]{| {#1} \rangle}

\begin{document}

\title{UNITARITY CUTTING RULES FOR HARD
PROCESSES ON NUCLEAR TARGETS
}

\author{W.~SCH\"AFER \\ \\
Institute for Nuclear Physics PAN \\
ul. Radzikowskiego 152 \\
31-342 Krak\'ow, Poland \\
E-mail: Wolfgang.Schafer@ifj.edu.pl }

\maketitle

\begin{abstract}
\noindent 
Unitarity cutting rules for the multiplicity of cut Pomerons, or topological 
cross sections have been obtained within the framework of 
nonlinear $k_\perp$ factorisation. Proper account for the color coupled channel
aspects of the problem leads to the emergence of two types of cut Pomerons.
This is illustrated on the example of topological cross sections in deep
inelastic scattering on a nucleus.
\end{abstract}



\markboth{\large \sl W. Sch\"afer 
\hspace*{2cm} HSQCD 2008} {\large \sl \hspace*{1cm} UNITARITY
CUTTING RULES}

\section{Introduction} 
It is a pleasure to present our results \cite{Cutting_Rules} 
here in Gatchina, where many of the fundamental ideas on
the use of unitarity cutting rules perhaps originated \cite{AGK}.

Consider, for example, (not too) small-$x$ deep inelastic 
scattering (DIS).
On the free nucleon target, the $q \bar q$ dipole will exchange
a single gluon with the target and the recoiling nucleon debris
will be in the color octet state. 
On a heavy nucleus target, multiple gluon exchanges are
enhanced by a new large parameter, the size of the nucleus.
At the parton level, the final state of a typical nondiffractive
event will now contain a certain number of color--excited nucleons. 
In the language of Pomeron exchanges, each color--excited
nucleon corresponds to a unitarity cut of a Pomeron exchanged
between the projectile $q\bar q$--dipole and the nucleus.

The partial cross sections associated with a fixed number 
of cut pomerons are called topological cross sections, and 
are important experimental observables. 
In the case of hard processes of interest to us, 
they contain information on the 
long--range rapidity correlations between forward or midrapidity 
jet/dijet production and multiproduction in the fragmentation region
of the target nucleus.
They are also closely related to the centrality of a collision,
a fundamental concept in nuclear interactions.

\section{Inelastic processes as transitions $a \to bc$ of beam partons}

\begin{figure}[!thb]
\begin{center}
\includegraphics[width=12.0cm,height=4cm]{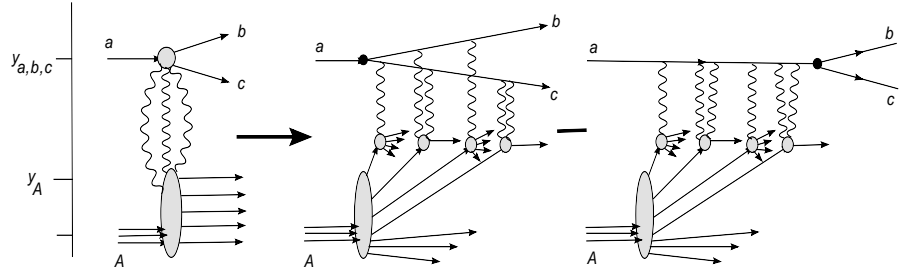}
\caption[*]{ 
A typical contribution to the excitation $a A \to bc X$ with
multiple color excitations of the nucleus. The amplitude receives
contributions from processes with interactions before and after
the virtual decay, which interfere destructively.}
\end{center}
\label{fig1}
\end{figure}

The $\gamma^* \to q \bar q$ transition of DIS is but the simplest 
example of a whole variety of processes involving the breakup
of a parton $a$ to its two--body Fock--space components $a \to bc$.
We will restrict ourselves to the situation where the $a \to bc$ transition
proceeds coherently over the whole nucleus. In DIS this means, for example,
$x \leq x_A = 1/2R_Am_p \approx 0.1 A^{-1/3}$.  
We can then make use of the fact, that the highly energetic partons
move along straight--line trajectories and their impact parameters $\bb_i$
are conserved during the interaction. When calculating the cross section, 
we can 
substitute the conjugated $S$--matrix of a parton by 
the $S$--matrix of an antiparton
 (see \cite{Slava} and references therein). 
Then, all we need are the $S$--matrices
for the forward scattering of fictitious multi--parton systems off the
nuclear target. They have been obtained for all cases of interest:
$\gamma^* \to q \bar q$ \cite{DIS}, which is of 
relevance for DIS and dijets/single jets
in the current fragmentation region, $q \to qg$ \cite{Quark_Gluon}, 
relevant for the production
of (di-)jets in $pA$--collisions at forward rapidities, $g \to gg$
\cite{Gluon_Gluon},
relevant for midrapidities, and $g \to Q \bar Q$
\cite{Nonuniversality}, for the production
of heavy quarks. 
The master formula for the dijet spectrum reads \cite{Single_jets}:
\begin{eqnarray}
&&\frac{d \sigma (a^* \to bc)}{dz_b d^2\bp_b d^2\bp_c} = \int 
{d^2\bb_b d^2\bb_c d^2\bb'_b d^2\bb'_c \over (2 \pi)^4} 
\exp[-i\bp_b(\bb_b-\bb_b')
- i \bp_c (\bb_c - \bb_c')] 
\nonumber \\
\times &&
\psi_{a \to bc}(z_b , \bb_b - \bb_c) 
\psi_{a \to bc}^*(z_b,\bb'_b - \bb'_c) 
\nonumber \\
\times &&\Big\{ 
S^{(4)}_{\bar{b}\bar{c}cb}(\bb'_b,\bb'_c,\bb_b,\bb_c)
+S^{(2)}_{\bar{a}a}(\bb'_a,\bb_a)
-S^{(3)}_{\bar{b}\bar{c}a}(\bb_a,\bb'_b,\bb'_c)
-S^{(3)}_{\bar{a}bc}(\bb'_a,\bb_b,\bb_c)
\Big\} \, .
\label{Master_eq}
\end{eqnarray}
Here $\psi_{a \to bc}(z_b , \bb_b - \bb_c)$ 
is the light cone wave function for the transition $a \to bc$,
spin/flavor etc. labels are suppressed. The fraction of $a$'s 
light--cone momentum carried by $b$ is $z_b$, and the transverse 
momenta of partons $b,c$ are $\bp_b,\bp_c$.
The multiparton $S$--matrices can be evaluated using the Glauber--Gribov
theory, for example the four--parton $S$--matrix will have the form
$S^{(4)}_{\bar{b}\bar{c}cb}(\{\bC \} ) 
= \exp[- \Sigma^{(4)}(\bC) T_A(\bb)/2]$,
where $\bC$ is a collective label for impact parameters, $\Sigma^{(4)}(\bC)$
is the color dipole (CD) cross section operator for the interaction
of the $bc\bar b \bar c$ system with the free nucleon target. It is an
operator in the space of possible color--singlet states
$\ket{R \bar R} = \ket{(bc)_R (\bar b' \bar c')_{\bar R} }$.

The momentum space formulation gives rise to the nonlinear $k_\perp$ 
factorisation \cite{DIS,Quark_Gluon,Gluon_Gluon,Nonuniversality,
Single_jets}, for short reviews, see \cite{Overview}.

To calculate the topological cross sections, we have to insert into eq.
(1) the multiparton $S$--matrices with a definite number 
of cut Pomerons. To this end, we decompose the free--nucleon CD operator
$\Sigma^{(4)}(\bC) = 
\Sigma^{(4)}_{el}(\bC) + \Sigma^{(4)}_{ex}(\bC)$
into an `elastic' part $\Sigma^{(4)}_{el}(\bC)$, which contains the 
contribution of color singlet exchange with a nucleon (the 
\emph{uncut} Pomeron),
and an `excitation' part $\Sigma^{(4)}_{ex}(\bC)$, which represents the 
(square of) color--octet single gluon exchange with a nucleon
(the \emph{cut} Pomeron). Then, an expansion of the nuclear $S$--matrix
in terms of $\Sigma^{(4)}_{ex}$ is precisely the sought--for expansion 
in terms of cut Pomerons. An expansion in terms of $\Sigma^{(4)}_{el}(\bC)$
would give rise to an expansion in terms of uncut Pomerons -- the sign
alternating multipomeron absorptive corrections. It is important to realize,
that the cut and uncut Pomeron parts of
$\Sigma^{(4)}$ separately are infrared sensitive, and depend on 
a nonperturbative parameter, the CD cross section for large
dipoles $\sigma_0$.

An important aspect of the color coupled--channel scattering is the emergence
of \emph{two types of cut Pomerons}. Indeed one has to distinguish
between the gluon exchanges which induce transitions between different
color multiplets $R_i \to R_j$ of the $bc$--system, and such color octet
exchanges which only rotate the $bc$ system within the same multiplet.

\section{Glauber--AGK vs. QCD}
As we lack the space, we skip here the discussion of topological cross 
sections for dijet observables, which follow from evaluating
eq.(\ref{Master_eq}). Instead, we present results for fully inclusive
topological cross sections in DIS, which illustrate the importance of a 
correct treatment of the color--coupled channel aspects.

Following early works on hadron--nucleus scattering \cite{GlauberAGK}, the 
profile function for the inelastic $q \bar{q}$-dipole-Nucleus interaction
could be expanded in manifestly positive terms:
\begin{eqnarray}
\Gamma^{inel} (\br, \bb) &=& 1 - \exp[- \sigma(\br) T(\bb)] 
= \exp[-\sigma(\br) T(\bb)]\sum_k {1 \over k!} [\sigma(\br) T(\bb)]^k 
\, .
\end{eqnarray}
Then one would interpret 
\begin{eqnarray}
\Gamma^{(k)}(\bb,\br) = { (\sigma(\br) T(\bb))^k \over k!} \, 
e^{-\sigma(\br) T(\bb)}
\end{eqnarray}
as the profile function for the $k$--cut Pomeron topological cross section, we
call it the `Glauber--AGK' prediction.
It is too good to be true -- a prediction of topological cross sections without
any soft parameter, given solely in terms of the CD cross section! Recall  
that the decomposition into cut and uncut Pomerons was infrared sensitive, and 
indeed this shows up in the explicit dependence of our result 
\cite{Cutting_Rules} on $\sigma_0$:
\begin{eqnarray}
\Gamma^{(k)}(\bb,\br) = \sigma(\br) T(\bb) \, \cdot \, w_{k-1}(2 \nu_A(\bb) )
\cdot {e^{-2 \nu_A(\bb)} \over \lambda^k} \gamma(k,\lambda) 
\, .
\nonumber
\end{eqnarray}
where
\begin{equation}
\lambda = 2 \nu_A(\bb) - \sigma(\br) T(\bb), \, \nu_A(\bb) = {1 \over 2} \sigma_0 T(\bb)  
\, , 
\gamma(k,\lambda) = \int_0^\lambda dt \,  t^{k-1} e^{-t}
\end{equation}
and $w_k(\nu_A) = \exp[-\nu_A] \nu_A^k / k!$. The dramatic difference between the 
single--channel 'Glauber--AGK' result and the novel QCD cutting rules is shown in
fig.(2), where we plot the profile function for $k$ cut Pomerons as a function
of impact parameter for a fairly large dipole of $r = 0.6$ fm, with a CD cros section
from \cite{EIC}. While multiple
cuts are nonnegligible also for the Glauber--AGK result, it suggests a strong hierachy
of topological cross sections. This is not at all borne out by our QCD cutting rules. 
In fig.(3) we show the $k$--cut pomeron contribution to the structure function $F_2$ on
a heavy nucleus as a function of impact parameter. 
The drastic difference to Glauber--AGK of our result is an 
interesting prediction for a possible future electron--ion collider
\cite{EIC_review}.

\begin{figure}[!thb]
\begin{center}
\includegraphics[width=6.0cm,height=6cm,angle=270]{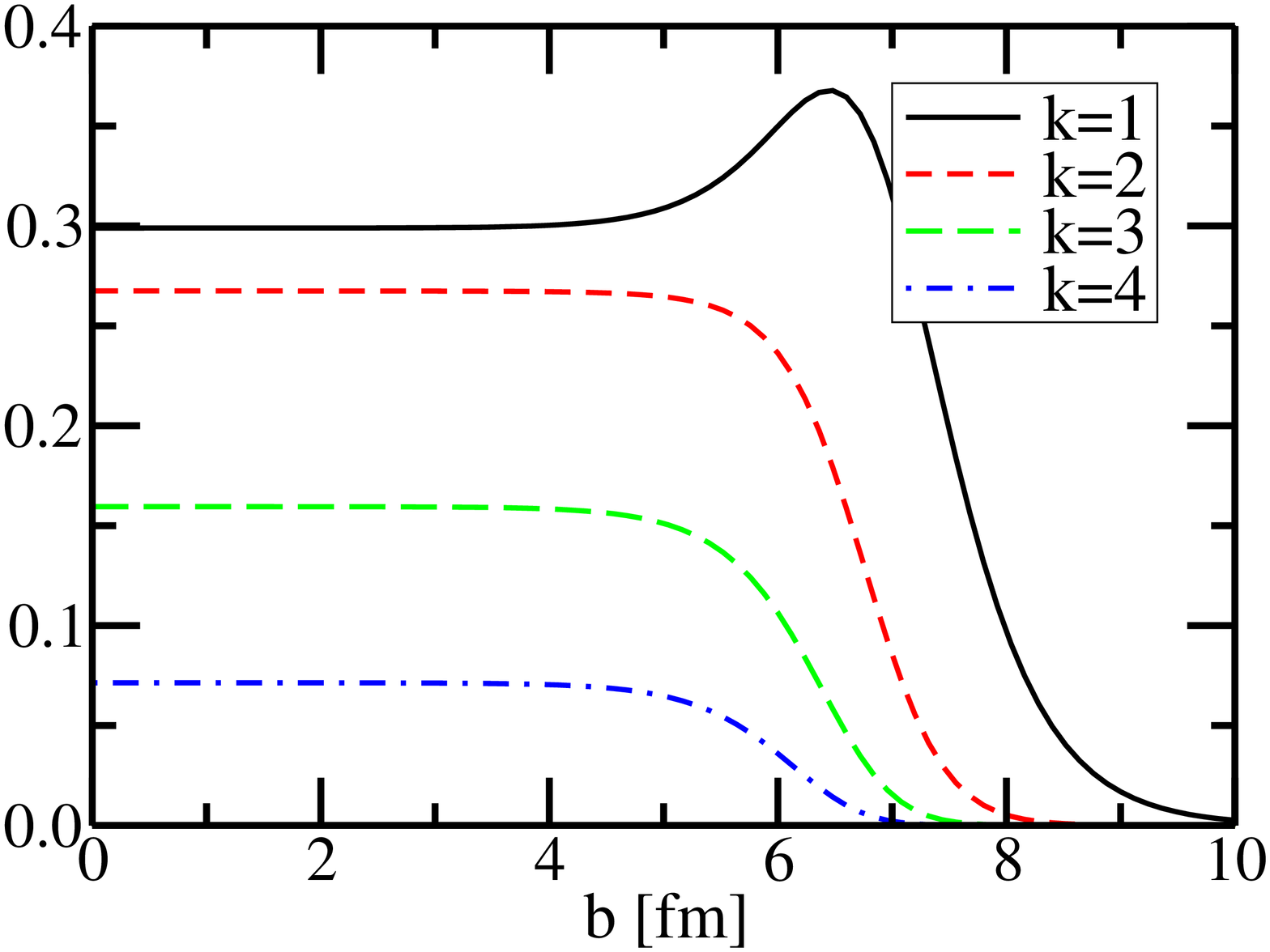}
\includegraphics[width=6.0cm,height=6cm,angle=270]{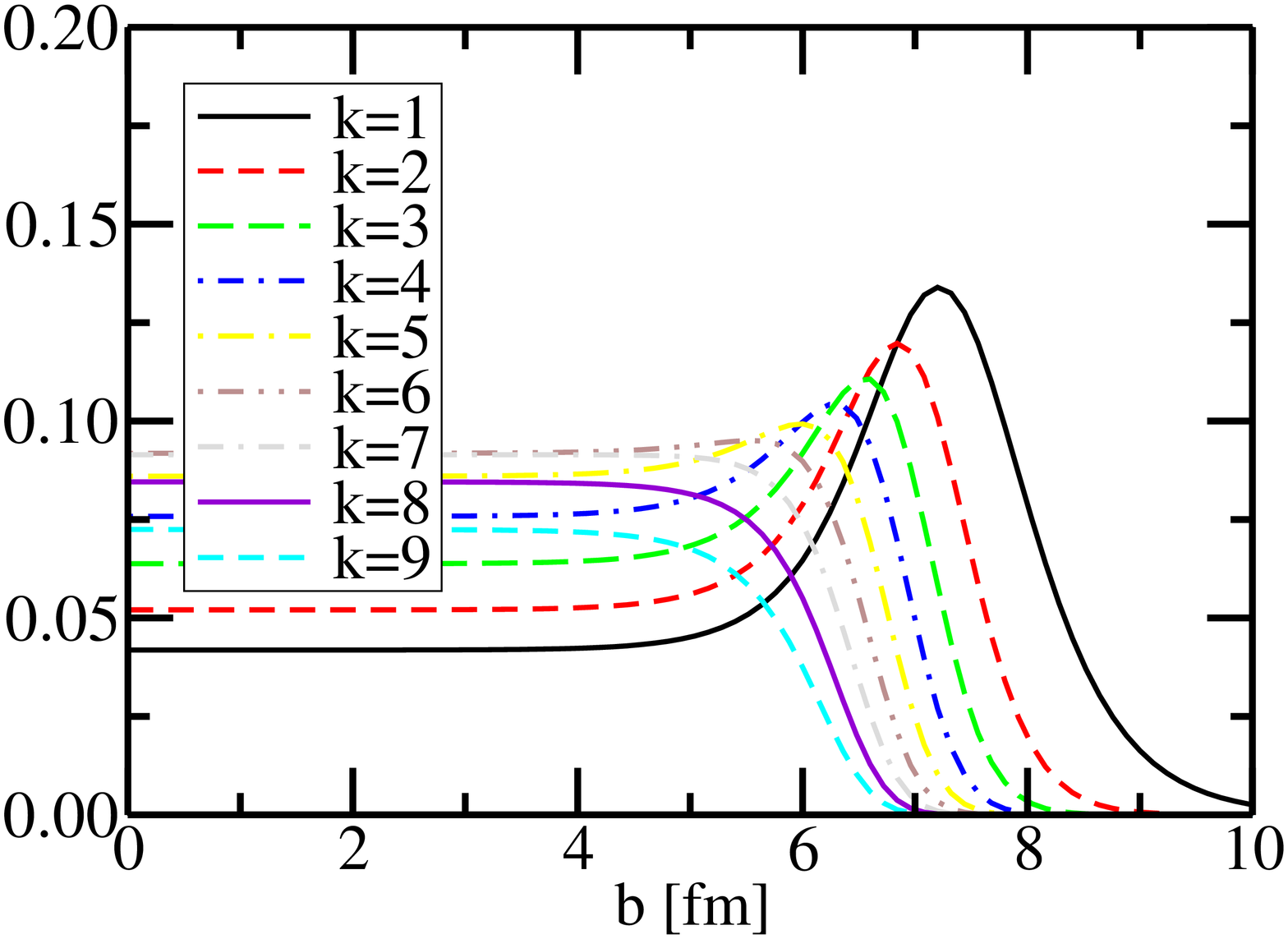}
\caption[*]{The profile function for $k$ cut Pomerons $\Gamma^{(k)}(\bb,\br)$
for $r=0.6$ fm, $x=0.01$, $A=208$. {\bf {Left:}} Glauber--AGK. {\bf{Right:}} 
QCD cutting rules.}
\end{center}
\end{figure}

\begin{figure}[!thb]
\begin{center}
\includegraphics[width=6.0cm,height=6cm,angle=270]{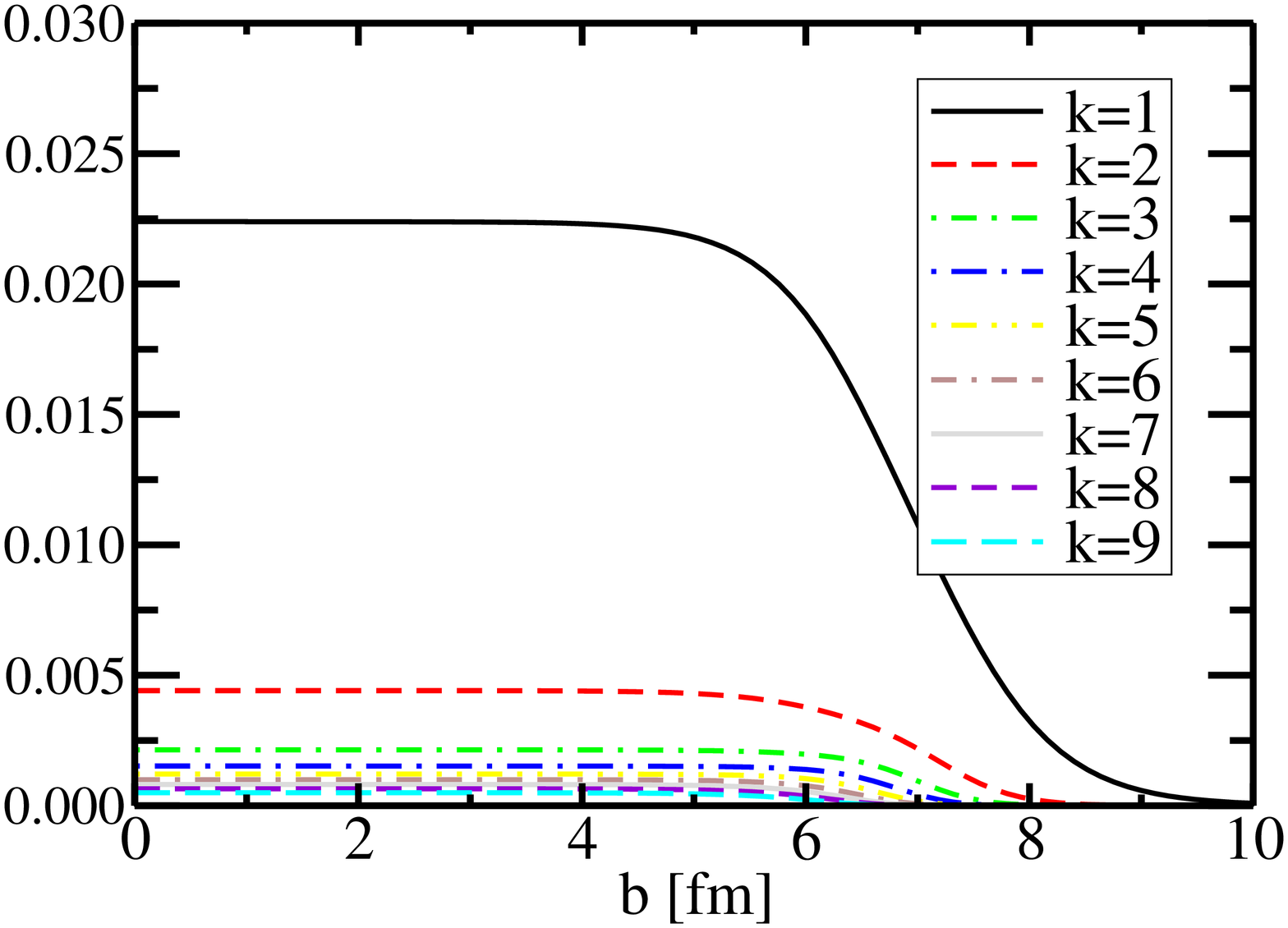}
\includegraphics[width=6.0cm,height=6cm,angle=270]{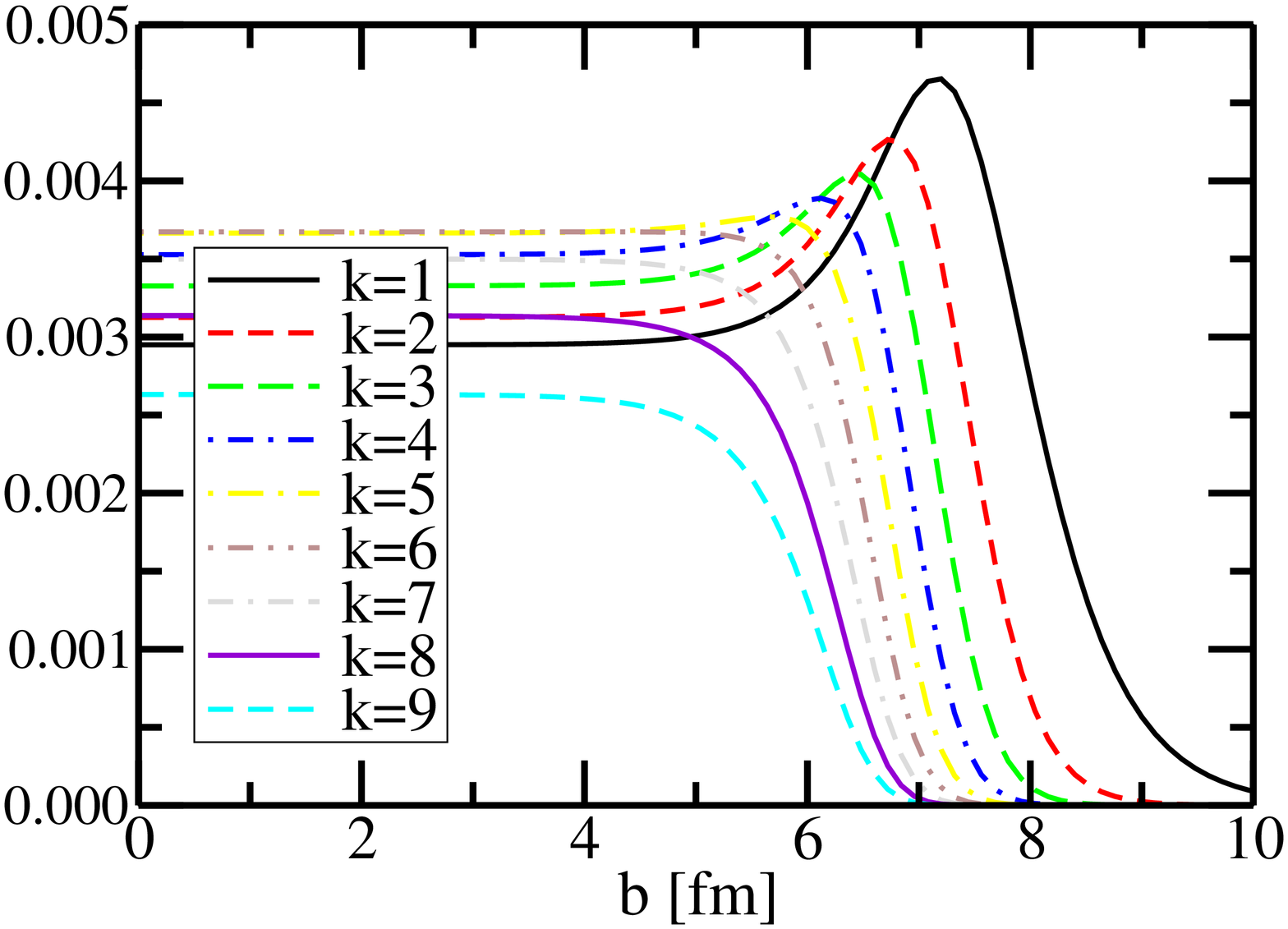}
\caption[*]{The $k$--cut Pomeron contribution $dF^{(k)}_2(x,Q^2)/d^2\bb$ to the DIS
structure function $F_2$ at $x=0.01$, $Q^2 = 10$ GeV$^2$, $A = 208$. 
{\bf{Left:}} Glauber--AGK, {\bf{Right:}} QCD--cutting rules.}
\end{center}
\end{figure}

\section*{Acknowledgements} 
This work was partially supported by the Polish Ministry of Science
and Higher Education (MNiSW) under contract 
1916/B/H03/2008/34.


\begin{thebibliography}{0}

\bibitem{Cutting_Rules}
  N.~N.~Nikolaev and W.~Sch\"afer,
  Phys.\ Rev.\  D {\bf 74} (2006) 074021.

\bibitem{AGK}
  V.~A.~Abramovsky, V.~N.~Gribov and O.~V.~Kancheli,
  Yad.\ Fiz.\  {\bf 18} (1973) 595
  [Sov.\ J.\ Nucl.\ Phys.\  {\bf 18} (1974) 308.

\bibitem{Slava}
  B.~G.~Zakharov,
  Nucl.\ Phys.\ Proc.\ Suppl.\  {\bf 146} (2005) 151
  [arXiv:hep-ph/0412117].


\bibitem{DIS}
  N.~N.~Nikolaev, W.~Sch\"afer, B.~G.~Zakharov and V.~R.~Zoller,
  J.\ Exp.\ Theor.\ Phys.\  {\bf 97} (2003) 441
  [Zh.\ Eksp.\ Teor.\ Fiz.\  {\bf 124} (2003) 491]


\bibitem{Quark_Gluon}
  N.~N.~Nikolaev, W.~Sch\"afer, B.~G.~Zakharov and V.~R.~Zoller,
  Phys.\ Rev.\  D {\bf 72} (2005) 034033.

\bibitem{Gluon_Gluon}
  N.~N.~Nikolaev, W.~Sch\"afer and B.~G.~Zakharov,
  Phys.\ Rev.\  D {\bf 72} (2005) 114018.

\bibitem{Nonuniversality}
  N.~N.~Nikolaev, W.~Sch\"afer and B.~G.~Zakharov,
  Phys.\ Rev.\ Lett.\  {\bf 95} (2005) 221803.

\bibitem{Single_jets}
  N.~N.~Nikolaev and W.~Sch\"afer,
  Phys.\ Rev.\  D {\bf 71} (2005) 014023.

\bibitem{Overview}
  W.~Sch\"afer,
  PoS {\bf DIFF2006} (2006) 040
  [arXiv:hep-ph/0611070];
  N.~N.~Nikolaev, W.~Sch\"afer, B.~G.~Zakharov and V.~R.~Zoller,
  JETP Lett.\  {\bf 82} (2005) 325
  [Pisma Zh.\ Eksp.\ Teor.\ Fiz.\  {\bf 82} (2005) 364].

\bibitem{GlauberAGK}
  A.~Capella and A.~Kaidalov,
  Nucl.\ Phys.\  B {\bf 111} (1976) 477;
  L.~Bertocchi and D.~Treleani,
  J.\ Phys.\ G {\bf 3} (1977) 147.

\bibitem{EIC}
  N.~N.~Nikolaev, W.~Sch\"afer, B.~G.~Zakharov and V.~R.~Zoller,
  JETP Lett.\  {\bf 84} (2007) 537.


\bibitem{EIC_review}
  A.~Deshpande, R.~Milner, R.~Venugopalan and W.~Vogelsang,
  Ann.\ Rev.\ Nucl.\ Part.\ Sci.\  {\bf 55} (2005) 165
  [arXiv:hep-ph/0506148].


\end{thebibliography}
\end{document}